\documentclass[useAMS,usenatbib]{mn2e}
\usepackage{aastex_hack}
\usepackage{graphicx}
\usepackage{amsmath}

\title[The VAMPIRES instrument]{The VAMPIRES instrument: Imaging the innermost regions of protoplanetary disks with polarimetric interferometry}

\author[B. Norris et al.]
{Barnaby Norris,$^{1}$ \thanks{E-mail: bnorris@physics.usyd.edu.au} 
Guillaume Schworer,$^{1,2}$ 
Peter Tuthill,$^{1}$ 
Nemanja Jovanovic,$^{3}$ 
 \newauthor
Olivier Guyon,$^{3}$ 
Paul Stewart$^{1}$ and 
Frantz Martinache$^{4}$\\
$^{1}$Sydney Institute for Astronomy, School of Physics, Physics Road, University of Sydney, N.S.W. 2006, Australia\\
$^{2}$LESIA, Observatoire de Paris, Section de Meudon, 5 place Jules Janssen, 92195, Muedon, Cedex\\
$^{3}$National Astronomical Observatory of Japan, Subaru Telescope, 650
N. A’Ohoku Place, Hilo, Hawaii 96720, U.S.A\\
$^{4}$Laboratoire Lagrange, CNRS UMR 7293, Observatoire de la Côte d’Azur, Bd de l’Observatoire, 06304 Nice, France\\
}

\begin{document}

\date{Accepted YYYY Month DD. Received YYYY Month DD; in original form YYYY Month DD}

\pagerange{\pageref{firstpage}--\pageref{lastpage}} \pubyear{2014}

\maketitle

\label{firstpage}

\begin{abstract}
Direct imaging of protoplanetary disks promises to provide key insight into the complex sequence of processes by which planets are formed. 
However imaging the innermost region of such disks (a zone critical to planet formation) is challenging for traditional observational techniques (such as near-IR imaging and coronagraphy) due to the relatively long wavelengths involved and the area occulted by the coronagraphic mask.  
Here we introduce a new instrument -- VAMPIRES -- which combines non-redundant aperture-masking interferometry with differential polarimetry to directly image this previously inaccessible innermost region. By using the polarisation of light scattered by dust in the disk to provide precise differential calibration of interferometric visibilities and closure phases, VAMPIRES allows direct imaging at and beyond the telescope diffraction limit. Integrated into the SCExAO system at the Subaru telescope, VAMPIRES operates at visible wavelengths (where polarisation is high) while allowing simultaneous infrared observations conducted by HICIAO. Here we describe the instrumental design and unique observing technique and present the results of the first on-sky commissioning observations, validating the excellent visibility and closure phase precision which are then used to project expected science performance metrics. 
\end{abstract}


\section{Introduction}
The mechanism by which planets are formed within circumstellar disks is a key question in current astronomy. Flattened, cool disks of gas and dust surround most low-mass stars for their first several millions of years of existence, during which they gradually dissipate via photo-evaporation, mass outflow, assimilation by the star, and condensation into planetisimals and eventually planets \citep{Williams2011}. 
Of particular interest are so-called transition-disks - protoplanetary disks exhibiting a partially evacuated gap, first identified via a distinctive dip in their infrared spectral energy distributions (e.g. \cite{Calvet2002}). An exciting possibility is that these gaps are indicative of planetary formation \citep{Bryden1999} and detailed observational characterisation is of particular importance. 

Although our understanding of the evolutionary processes involved is incomplete, in recent times spatially resolved observations have provided great insight into the structure and evolution of such protoplanetary disks, including the gaps, knots and other density modulations that provide evidence of planetary formation. Sub-millimetre observations have resolved the inner cavities of transition disks \citep{Andrews2011}, as has long-baseline optical interferometry \citep{Olofsson2011}. Furthermore, infrared observations using coronagraphy and polarimetry have revealed fine substructure within protoplanetary disks. Spiral arms and complicated asymmetrical structure has been imaged using these techniques in several disks, such as those surrounding AB~Aurigae \citep{Fukagawa2004, Hashimoto2011} and MWC~758 \citep{Grady2013}.

While such infrared observations are extremely productive, the imaging of the innermost region (within $\sim$100 milliarcseconds) is highly challenging, due to the high contrast and angular resolution required as well as the area occulted by the coronagraphic mask in traditional coronagraphs.
However this inner region is critical for the understanding of disk structure and planet formation \citep{Williams2011}. The main disk inner rim, the partially evacuated cavity (such as that characteristic of transition disks) and an inner-disk may all lie within this region. 
Here we introduce a new instrument, the Visible Aperture-Masking Polarimetric Interferometer for Resolving Exoplanetary Signatures (VAMPIRES), specifically designed to directly observe this key inner region. Using a unique combination of aperture-masking interferometry and differential polarimetry, VAMPIRES will directly image the structure of the inner disk region in scattered starlight at visible wavelengths, revealing rim structure, disk geometry, and asymmetric density perturbations betraying the presence of an accreting planetesimal. Furthermore polarimetric measurements reveal the distribution of dust grain sizes and species. Integrated into the SCExAO (Subaru Coronagraphic Extreme Adaptive Optics) system \citep{Guyon2011, Jovanovic2013} at the Subaru 8~m telescope, it will enable observations at the telescope diffraction limit. The observational parameter space explored by VAMPIRES is distinct from -- and complementary to -- that of a coronagraphic polarimetric imager; it has no limitation on inner working angle (limited only by its resolving power, around 10~mas) and offers diffraction-limited imaging of this inner region while being robust against seeing and imperfect AO correction.

The remainder of this section will outline the technical background to the technique employed by VAMPIRES. Section~\ref{TheInstrument} will describe the instrument itself. Section~\ref{datareduction} describes the unique triple-differential interferometric calibration used by VAMPIRES. The results from the first on-sky observations are presented in Section~\ref{onsky}. A qualitative demonstration of the expected data from science observations is given in Section~\ref{sec_simdata} and a summary is included in Section~\ref{Summary}. Appendix \ref{IPCal} explains the calibration procedure used for residual instrumental polarisation not removed during the differential process. 

\subsection{Non-redundant aperture-masking}
\label{NRM}
Non-redundant aperture-masking \citep{Readhead1988,Tuthill2000}, segments the pupil of a single telescope into a number of sub-apertures by placing an opaque metal mask with a carefully designed pattern of holes at a pupil plane upstream of an imaging instrument. This causes a diffraction pattern to form on the detector, with each set of fringes corresponding to a particular pair of sub-apertures (holes) in the aperture mask. Thus each pair of sub-apertures forms a baseline of a Fizeau interferometer, with the visibilities and phases recovered by Fourier analysis of the diffraction pattern.

The key to the technique's performance is that the layout of sub-apertures must be \emph{non-redundant} -- that is, the vector separation of every hole pair must be unique. This allows for the noise-process arising from seeing to be largely eliminated (see \cite{Readhead1988} for more detail).
The power (or squared visibility) recorded for fringes on any baseline is intrinsically robust against phase errors caused by seeing, and this observable alone provides powerful constraint for diffraction limited imaging from terrestrial telescopes. However visibilities do not preserve any of the phase information of the image. Ideally the phases recorded for each baseline would also be utilised, however these are completely dominated by random error from seeing. However an alternate observable -- the closure phase \citep{Baldwin1986} -- can be derived, which in the limit of small sub-apertures and short exposures, is immune to the effects of seeing. When starlight entering each sub-aperture is corrupted by a random phase error, then by taking the sum of phases around three baselines forming a closing triangle, phase errors cancel out and the resultant observable, the closure phase, is a function only of the source intensity distribution. The use of non-redundant masking along with these two observables has been highly successful in producing diffraction-limited, high-contrast images of such targets as stellar surfaces and atmospheres of evolved stars \citep{Haniff1987, Tuthill1999, Woodruff2008}, dusty plumes surrounding Wolf-Rayet stars \citep{Tuthill1999b} and, most recently, protoplanetary disks \citep{Eisner2009,Huelamo2011} and even suspected sub-stellar companions undergoing formation within \citep{Arnold2012,Kraus2012}. 

Both visibilities and closure phases are subject to systematic errors arising from imperfections and instabilities in the instrumental point-spread function (PSF). To combat this, the conventional approach in interferometry is to interleave observations of the science targets with observations of a calibrator star, usually an unresolved point-source (or sometimes an object with well known structure). The calibrated visibilities are formed from the ratio of the science target's visibilities to those of the calibrator star, and the calibrated closure phases are the difference between those of the science and calibrator stars. The success of this calibration assumes that aberrations encountered by the calibrator star are a good statistical representation of those for the science target. This can be a poor assumption if the two are at different air-masses or if conditions change between observations.

Current masking interferometry programs usually deploy the aperture-mask in a beam corrected by an adaptive optics system. For such experiments \citep{Tuthill2006}, the AO system acts as a fringe-tracker, stabilising the fringe phase so that visibility information is preserved for exposure times longer than the atmospheric coherence time (a fraction of a second). 
The precision calibration provided by non-redundant masking then allows the PSF of the AO-corrected imaging system to be accurately calibrated.

\subsection{Astronomical polarimetry}
Polarimetry and polarimetric imaging are well established techniques in astronomy. Earlier polarimetric studies measured the overall polarisation of light from a star using a polarimeter with a single pixel detector such as a photomultiplier tube (e.g. \cite{Hall1950, Hiltner1956, Gehrels1960, Mathewson1970}). These instruments had polarisation precisions as good as 0.01\% when used over narrow bands, with subsequent developments extending this precision to wide bandwidths (e.g. \cite{Tinbergen1973}.) Later, imaging polarimeters were developed that could measure the polarisation of a spatially resolved field, using both photographic \citep{Woltjer1957} and electronic \citep{Scarrott1983} methods. General-purpose imaging instruments on 8~m telescopes such as NACO on the VLT and AO188/HICIAO on the Subaru telescope offer polarisation modes, wherein the emphasis is on polarimetric differential imaging (PDI) -- using the difference in polarisation between starlight and scattered light to solve the contrast-ratio problem of imaging circumstellar disks -- rather than high polarisation precision (which is of order 1\% for NACO \citep{Witzel2011}). Newly developed polarimeters such as SPHERE/ZIMPOL \citep{Thalmann2008} promise sensitivities on the order of order 1 part in $10^5$.

As in conventional imaging, all these polarimeters operate by forming an image of the science target on an array detector (such as a CCD), but with the addition of polarisation optics which convert the polarisation properties of the signal into intensity variations. In the simplest case this could be an analyser (linear polariser), but in practice more complex systems are used to achieve the desired sensitivity and mitigate the effects of systematic errors resulting from instrumental polarisation. Dual beam methods (where orthogonal polarisations are measured simultaneously using a polarising beam-splitter) are often used, 
originally designed to improve polarimetric precision by removing the effect of seeing \citep{Hiltner1951}. 
A related technique, channel switching (where polarisations are switched upstream with a rotating half-wave plate or similar) is often used in combination with dual beam methods \citep{Appenzeller1967}. Here, the channel-switching device is placed as far upstream as possible (before polarisation-modifying elements just as mirrors and filters) allowing these systematic errors to be cancelled. Using either or both of these techniques, the differences or ratios of the intensities in the channels can be used to mitigate the effects of seeing and of instrumental polarisation \citep{Bagnulo2009}.

High precision instruments may use fast polarisation switching (using a liquid crystal device or similar) to modulate the polarisation on timescales equal to or faster than atmospheric turbulence \citep{Tinbergen1973}. Initially restricted to use with single-pixel detectors, advances have subsequently allowed this technique to be used with array detectors via some from of de-modulation at the detector, using an optical demodulator \citep{Stenflo1985} or an on-detector charge-shuffling technique \citep{Bazzon2012, Thalmann2008}. 
A particular challenge is the changing instrumental polarisation at the Nasmyth focus of modern alt-az telescopes due to the changing angle of the M3 mirror with respect to the Nasmyth platform - a challenge that can be addressed by using a rotating compensator plate system \citep{Tinbergen2007}. In all cases, proper characterisation of and calibration for instrumental polarisation is necessary.

In terms of exoplanet and protoplanetary disk studies, single-pixel polarimeters rely on detecting the small perturbations made to the overall polarisation of the stellar system by the presence of a planet or other asymmetric body. They therefore need to be extremely sensitive to small polarisations; even systems with a close-in giant planet are expected to exhibit polarisations of only $10^{-5}$ or so \citep{Seager2000}. However if the planet or disk is spatially resolved (as in an imaging polarimeter) then a much lower polarimetric sensitivity is required, assuming the surface polarisation of the object is high. This methodology has been used successfully in the imaging of protoplanetary disks and the substructure therein, including around AB Aurigae \citep{Perrin2009}, HD100546 \citep{Quanz2011} and HD142527 \citep{Avenhaus2014}, in which fractional (surface) polarisations from several percent up to several tens of percent were observed in the near-IR. These adaptive-optics observations provided imaging of the disk to within an inner working angle of (in the best case) 100~mas from the star.

\subsection{A new approach: polarimetric differential non-redundant masking}
The VAMPIRES instrument is based upon a recent extension to the aperture masking technique: \emph{polarimetric} non-redundant masking, initially demonstrated using the NACO instrument on the VLT to image the dust shells around AGB stars \citep{Norris2012}. This technique -- and the VAMPIRES instrument -- leverage the aforementioned switching and modulation methods from conventional polarimetry and recast them in an interferometric context.
In contrast to proposed methods for full Stokes optical interferometry \citep{Elias2001}, we take a different, differential approach. 
Rather than calibrating the instrumental transfer function using a nearby reference star, this technique instead records two orthogonal polarisations of the science target simultaneously.
Differential polarised observables can then be extracted by calibrating these observations against one another.
In practice, extreme care must be taken to avoid systematic errors which would otherwise overwhelm the science signal - see Sections~\ref{TheInstrument} and \ref{datareduction}. 
This technique offers three distinct advantages in imaging circumstellar regions which emit even a modest degree of polarised flux. 

Firstly, it allows direct observations of polarised circumstellar structures at very small separations from the star, at contrasts which are unachievable with conventional non-redundant masking. Dusty structures such as mass-loss shells and circumstellar disks scatter unpolarised light from their host star, a process which yields a polarised signal. The differential polarimetric observables produced by this technique describe the spatially-resolved polarised intensity distribution, allowing faint, polarised structures to be clearly imaged in isolation despite the immediate proximity of the very bright (but unpolarised) stellar photosphere.

Secondly, imaging data for the two orthogonal polarisations are recorded simultaneously, as opposed to many minutes apart for more conventional stellar interferometry (the time taken to slew from science target to reference star).
For the interferometric calibration, this is a profound difference. 
Rather than attempting calibration of the statistical properties of the telescope-atmosphere transfer function, starlight recorded simultaneously traversed identical optical paths (with the exception of a small instrumental leg immediately prior to the detector discussed later).
Exact, frame-by-frame calibration then makes it possible to eliminate errors arising from temporal-variation such as seeing, imperfect AO correction, changing airmass and flexure of the optics. 

Thirdly, the polarimetric data produced can reveal important information about the scattering medium. By using multi-wavelength studies of the degree of fractional polarisation, quantities such as dust grain size and even chemical makeup can be accurately constrained. 

It is also important to compare polarimetric aperture masking to conventional imaging polarimetry. The key distinction in technique is that the differential calibration - performed by subtraction or division of intensities in conventional polarimetry - is instead performed in the Fourier domain by the calibration of interferometric visibilities. The goal of this technique is analogous to that of polarimetric differential imaging -- to exploit the difference in polarisation between a star and the surrounding circumstellar dust (such as a protoplanetary disk) to overcome the contrast ratio problem and directly image the circumstellar region. 

However, the parameter space in which polarimetric aperture-masking operates is quite distinct from that of conventional polarimetric imaging / PDI. High contrast, high angular-resolution imaging polarimeters generally employ a coronagraph to block out the light from the star to help achieve the desired contrast ratio between star and disk. This technique is thus limited by the coronagraph's inner working angle, with the two newest instruments - SPHERE and GPI -  having an inner working angle of $\sim$100~mas and $\sim$200~mas respectively \citep{Martinez2010}. This puts the innermost regions of circumstellar disks -- critical for proper understanding of the planetary formation process -- out of reach for conventional imaging polarimeters. A polarimetric aperture-masking instrument such as VAMPIRES, however, has an effective inner working angle limited only by its resolving power, approximately 10~mas (with a field of view of only $\sim$300~mas). Additionally, the spatial resolution of an imaging polarimeter is limited by the performance of its adaptive optics system, while aperture masking can provide diffraction-limited performance even with high levels of wavefront error. Thus this technique can explore a unique parameter space, unreachable with - but complementary to - conventional imaging or imaging polarimetry. 

In the comparison between conventional polarimetry and polarimetric aperture masking it is also important to note that the residual systematic errors in the differential visibilities (as shown in Section  \ref{onskydiff}) do not represent the absolute polarisation precision of the instrument. Rather, they arise from the difficulties in precisely calibrating visibilities common to all optical interferometry. Conventional aperture masking exhibits errors in calibrated visibilities of order 5\%-10\% \citep{Monnier2004}, while the polarimetric technique described here produces errors more than an order of magnitude better than this (see Section \ref{onskydiff}).

Polarimetric non-redundant masking offers one other advantage over a conventional imaging polarimeter: the Fourier transform of each interferogram is normalised with respect to the zero-baseline power, so the resulting visibilities are not a function of the total flux in the channel. This means that the derived observables are intrinsically immune to many forms of instrumental polarisation systematic error, such as differing levels of transmission through the two channels of the polarising optics. This automatically sidesteps a large fraction of instrumental challenges arising in more orthodox polarimeters, with the remainder being calibrated as described in Appendix~\ref{IPCal}.

\section{The VAMPIRES instrument}
\label{TheInstrument}
VAMPIRES is a purpose designed instrument for performing polarimetric differential non-redundant masking. It is integrated into the SCExAO Extreme-AO system at the 8~m Subaru telescope, first tested on-sky in July 2013. The concept grew from experience with the SAMPol mode commissioned by our group on the NACO instrument at the VLT \citep{Norris2012} which enables near-infrared polarimetric masking interferometry. However the new purpose-built VAMPIRES instrument offers a number of major advantages. While SAMPol relies on a strategy of slow-switching (every few minutes) of a half-wave plate to swap polarisation channels and thereby calibrate non-common path error, VAMPIRES features a fast-switching liquid-crystal variable retarder (LCVR) which allows channel switching every $\sim$ 10 milliseconds. This is part of a three-tiered differential calibration scheme that demonstrated on-sky polarised visibilities calibrated to a precision of 1 part in $10^{3}$ and polarised closure-phases calibrated to a fraction of a degree, as detailed in Section~\ref{onsky}.

A Wollaston prism (first tier) provides simultaneous measurements of orthogonal polarisations, allowing precise calibration of all time-varying components (e.g. due to seeing) but introduces non-common path errors. The fast-switching LCVR (second tier) works in concert, calibrating non-common path errors and variability on timescales longer than the switch time. Finally a half-wave plate (third tier) which is placed further upstream calibrates out residual instrumental polarisation and non-common path error of all optics downstream, particularly the non-ideal performance (e.g. chromaticity) of the LCVR. The entire instrument has been designed with precision instrumental polarisation calibration in mind, incorporating a polarisation state injector and rotating quarter-wave plate for instrumental characterisation. 

Whereas  NACO/SAMPol operates in the near-infrared J -- K bands, the visible/IR region (from 0.6 $\mu$m to 1.0$\mu$m) explored by VAMPIRES exploits the fact that the degree of polarisation induced by scattering is a strongly rising function with shorter wavelengths. 
At the limit of small particle sizes, this can be seen in the $\lambda^{-4}$ term in Rayleigh scattering. 
Moreover, at shorter wavelengths the dependence of scattering cross-section on particle-size (for Mie scattering) is more pronounced, allowing greater discrimination between grain sizes. 
Stronger scattered-light signals yield more benign contrast ratios between circumstellar material and bright unpolarised host stars, allowing a wider range of astrophysical targets to be observed. 
Furthermore, wavelengths shorter than 1 $\mu$m are favoured from the perspective of detector technology as sensitive, relatively inexpensive EMCCD cameras are available providing rapid-exposure frames that are essentially free of readout noise.

VAMPIRES operates in a `hitch-hiker' mode, taking data concurrently with infrared observations. The SCExAO system delivers wavelengths longer than 1~$\mu$m to its coronagraphs and IR science instruments (such as the HICIAO camera). Meanwhile, wavelengths shorter than 1~$\mu$m are routed to a second `visible' optical bench where the available band is further divided between SCExAO's pyramid wavefront-sensor, VAMPIRES, and potentially other instruments. Therefore VAMPIRES offers the opportunity to perform polarimetric aperture masking observations simultaneously with HICIAO imaging during observations of protoplanetary disks and other suitable targets, delivering a supporting data stream without cost in additional observing time.

The parameter space explored by VAMPIRES is highly complimentary to that of coronagraphs. VAMPIRES can deliver extremely high angular resolutions (of order 10~mas) over a limited field-of-view ($\sim$300\,mas depending of choice of mask and wavelength). Thus the {\it outer working angle} of VAMPIRES complements the {\it inner working angle} of traditional coronagraphs, which typically have occulting spots of a few hundred milliarcseconds \citep{Grady2013}. VAMPIRES’ small inner working angle (effectively defined by the telescope diffraction limit) is in line with SCExAO’s core mission of small inner-working angle wavefront control and coronagraphy, as implemented by its PIAA coronagraph \citep{Guyon2010} (whose inner working angle is as small as 50 milliarcseconds in the near-infrared).
This inner region also plays into the nature of expected astrophysical signals, in which scattered starlight is expected to exhibit a strong fall-off with disk radius.

Although the primary design driver for VAMPIRES is the differential polarised masking mode, the instrument does allow a more versatile set of observational configurations.
For example, it is entirely possible to treat VAMPIRES as a standard masking interferometer and employ a separate PSF reference star (as in Section~\ref{NRM}).
This would then deliver standard complex visibility imaging data (rather than polarisation-differential data), enabling normal interferometry science to be performed and sacrificing the exceptional differential calibration.
The outcome of operational testing of this mode is briefly discussed in Section~\ref{non-pol-nrm}.
An even more straightforward operational mode is to employ VAMPIRES as an imaging polarimeter. 
Reconfiguring so that the mask is removed from the beam, and an image-plane field stop deployed to prevent polarised-field overlap, then rapid exposure differential imaging polarimetry data can be recorded.
However the focus of the present paper is on the novel interferometric polarimetry and further discussion of these more orthodox strategies is limited. 

\subsection{Instrument Design Description}

\begin{figure*}
    \centering
    \includegraphics[width=1.0\textwidth]{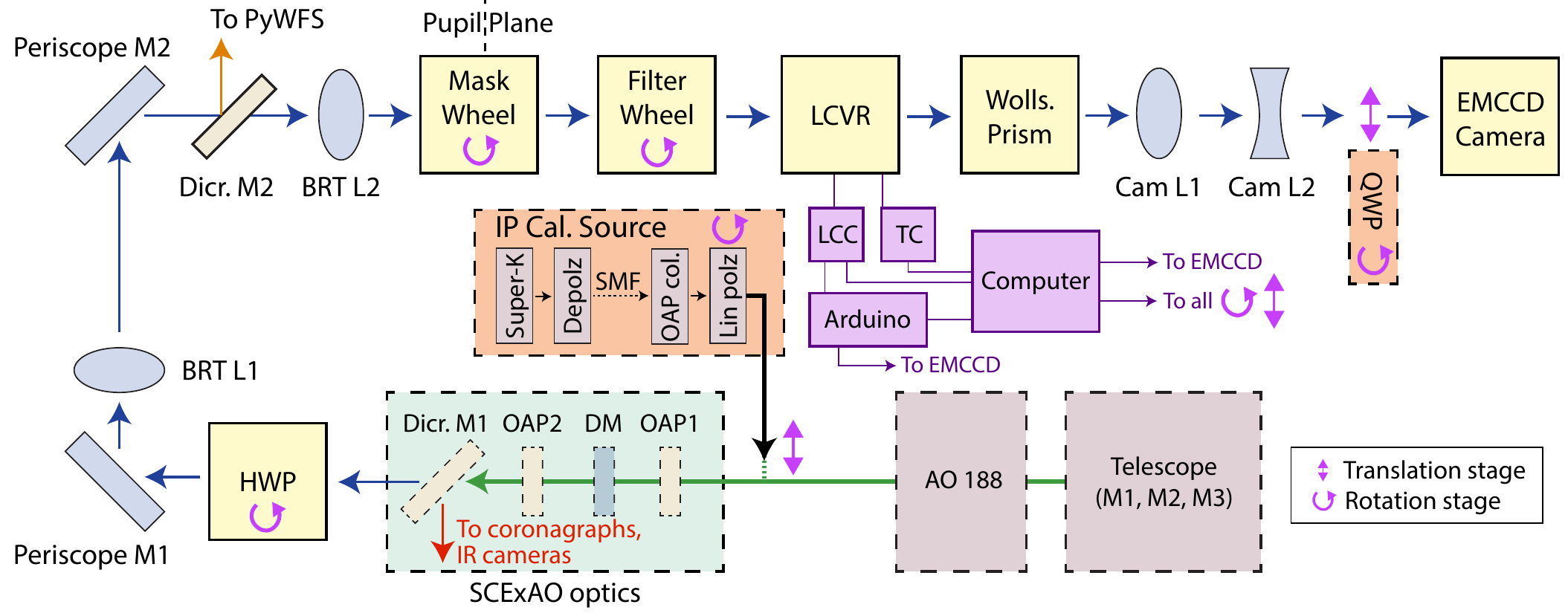}
    \caption{A schematic diagram of VAMPIRES as configured on-sky in July 2013, with all items relevant to the VAMPIRES beam train shown. Operation of each subsystem is described in the text. Abbreviations: M - Mirror; L -Lens; OAP - Off Axis Parabola; DM - Deformable Mirror; Dicr.M - Dichroic Mirror; HWP - Half-wave plate; BRT - Beam Reducing Telescope; LCVR - Liquid Crystal Variable Retarder; LCC - LCVR Controller; TC - Temperature Controller; Cam - Camera; QWP - Quarter-Wave Plate; Depolz - Depolariser; OAP col. - OAP Collimator; Lin polz - Linear polariser. In an alternative configuration, the half-wave plate can be replaced with a pair of quarter-wave plates to allow birefringence to be corrected as needed}
    \label{fig_schematic}
\end{figure*}

\begin{figure*}
    \centering
    \includegraphics[width=0.8\textwidth]{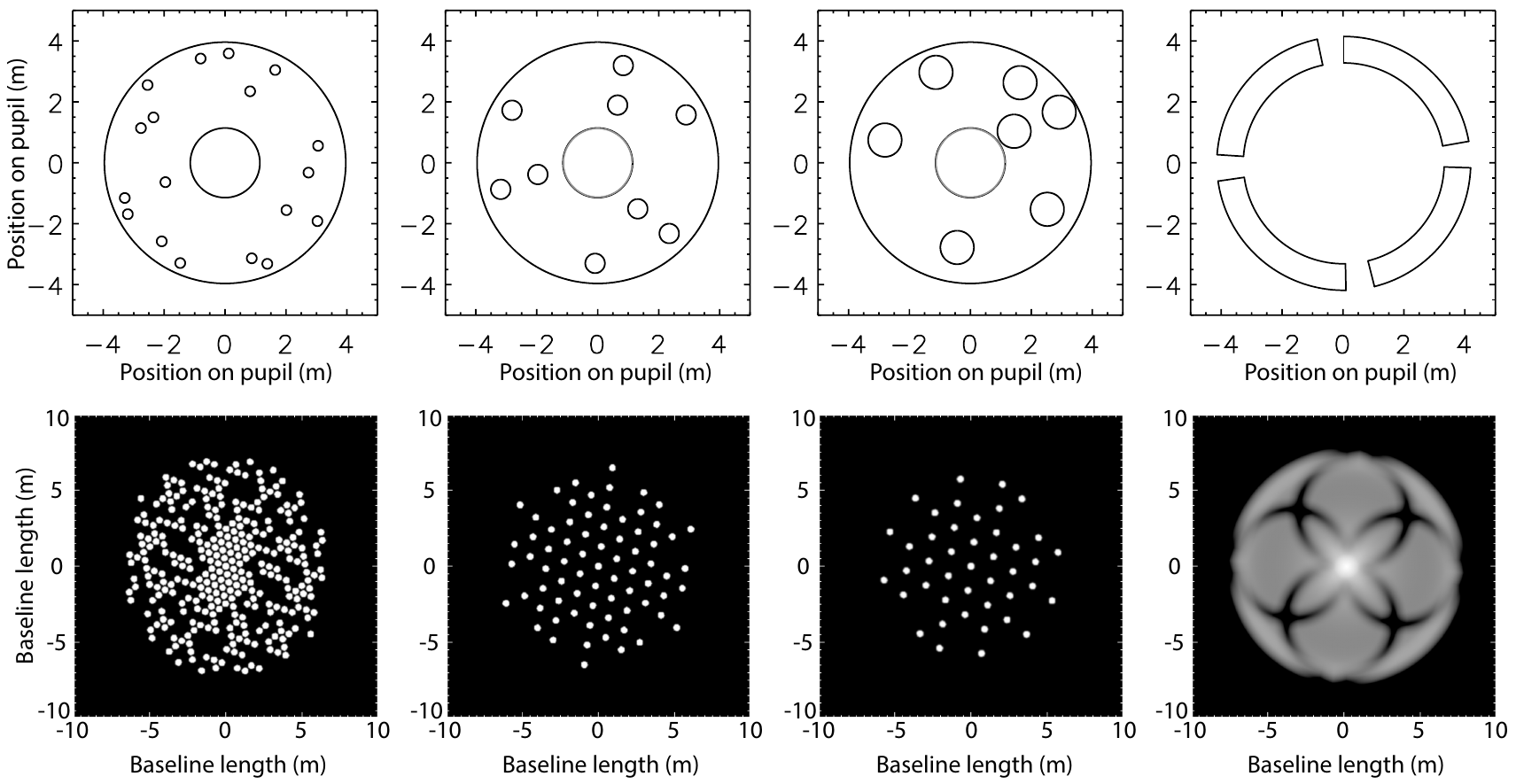}
    \caption{The non-redundant aperture mask designs installed in VAMPIRES (top) and their corresponding Fourier coverage (bottom). Masks with a greater number of holes boast better Fourier coverage at the expense of throughput. Masks are all non-redundant (the vector separation of all hole pairs is unique), with the exception of the annulus mask (right). This is doubly-redundant, sacrificing non-redundancy for high throughout and full Fourier coverage. The gaps in Fourier coverage for this mask are due to the missing portions of the annulus needed to screen out the secondary-mirror supports (spiders).}
    \label{fig_apmasks}
\end{figure*}

A schematic diagram of the instrument is given in Figure~\ref{fig_schematic}. VAMPIRES is integrated into the SCExAO extreme adaptive-optics system which consists of two optical tables mounted one above the other, with the lower bench dedicated mainly to the near-infrared channel and the upper bench dedicated to the visible. The beam path from the telescope and AO\,188 system passes into SCExAO, where high-order AO correction takes place. Wavelengths shorter than 1 $\mu$m are split off via a long-pass dichroic mirror and sent to the top `visible' bench via a periscope and beam-reducing telescope. VAMPIRES' half-wave plate (HWP) is mounted in a Newport CONEX-AG-PR100P piezo-motor rotation stage which is positioned in the reflected beam immediately after the dichroic mirror and prior to the periscope mirror. This places it as far upstream as possible, sampling the greatest possible number of optics which could potentially induce instrumental polarisation and therefore delivering the best possible calibration via channel switching (see Section~\ref{datareduction}). 
The HWP is a custom-designed achromatic waveplate manufactured by Casix, with $\lambda/2$ retardance from 600~nm to 1000~nm. An alternative configuration is also being investigated wherein the HWP is replaced by two quarter-wave plates (QWP). In this case, moving the QWPs together rotates the polarisation as per the HWP, but moving one QWP relative to the other allows birefringence in the system to be compensated for.

The beam diameter at the HWP is 18\,mm, but this reduces to 7.2\,mm as it travels to the top bench through a periscope that also contains a beam-reducing telescope (BRT) which consists of a Thorlabs AC508-500-B lens and AC508-200-B lens. A band-pass dichroic mirror is placed at the focus of the BRT to divert part of the spectrum to SCExAO's pyramid wavefront sensor (PyWFS). This dichroic can be exchanged under automated control to select the band directed to the PyWFS, and hence also the complementary band entering VAMPIRES. The BRT re-images the telescope pupil onto the non-redundant aperture-mask. The aperture masks available are shown in Figure~\ref{fig_apmasks}, and are located in a custom wheel (mounted on a CONEX-AG-PR100P rotation stage) allowing masks to be changed at will. Factors governing choice of mask can be complicated, but often boil down to achieving the best balance between Fourier coverage and throughput for a given science target. Following the mask, a wheel populated with filters ranging from 600~nm to 800~nm with bandwidths of 50~nm is used to select the observing band.

Next the beam encounters the liquid crystal variable retarder (LCVR), a Thorlabs LCC1111T-B device controlled by a Thorlabs LCC25 liquid crystal controller. The LCVR acts as variable wave-plate, switching its retardance between 30\,nm and $\lambda/2$ depending on the applied voltage. This switching takes place rapidly, with a rise time of $\sim$10~ms and fall time of $\sim$250~$\mu$s. The retardance is also a function of temperature, and specified values are required for successful calibration. Therefore the temperature of the device is actively controlled (using a Thorlabs TC200 temperature controller) to a temperature 45$^\circ$~C (higher temperatures allow faster switching times). The two orthogonal polarisations are then separated using a custom designed Wollaston prism (manufactured by Altechna), and focused onto the detector using a pair of lenses (Thorlabs AC254-200-B and ACN254-075-B) in a telephoto configuration. The detector is an Andor Ixon 897 Ultra EMCCD camera. 

The system is controlled by custom software written in Matlab on a Linux-based Intel i7 computer. This software allows control and scripting of all aspects of data acquisition (interfacing with the EMCCD camera, Arduino, and LCVR controller) as well as reconfigurable optomechanics.
The software allows control of all functions via single graphical user interface, allowing simple remote operation and permitting the automation of lengthy calibration procedures. 

A separate subsystem allows off-line characterisation of the LCVR and calibration of residual polarisation errors throughout the internal instrumental optical path. A calibration source (common to SCExAO and VAMPIRES) can remotely inject a simulated telescope beam into SCExAO with a specified polarisation state. 
This consists of a Super-continuum source
coupled into a single-mode fibre via a Thorlabs DPU-25-B achromatic depolariser to remove any prior intrinsic polarisation. The light is then collimated and passed through an achromatic linear polariser (Thorlabs LPVIS100-MP) mounted in a CONEX-AG-PR100P rotation stage to allow linear polarised light at any specified orientation to be injected. A quarter-wave plate (Thorlabs AQWP10M-980), also in a rotation mount, is also moved into the beam allowing a complete Mueller matrix of the instrument to be constructed as described in Appendix \ref{IPCal}.

As described in Section~\ref{datareduction}, VAMPIRES employs three tiers of differential calibration, one of which is rapid channel switching using the LCVR. This requires alternate frames acquired by the camera to have the incident beam polarisation rotated by 90$^\circ$, thus swapping the state probed by the two Wollaston channels. Due to the high acquisition rate and short integration times used ($\sim$17~ms in order to maintain high visibilities despite residual seeing after AO correction) it was not possible to directly control the LCVR switching and camera exposures using the computer, due to the non-realtime operating system and variable USB latency. Instead, the timing signals were generated by a dedicated Arduino Uno microcontroller. When a data acquisition cycle is initiated at the computer, the desired timing parameters and commands are sent to the Arduino. The Arduino sends the appropriate timing signals based on its internal clock (via 5V TTL pulses) to the LCVR controller and camera. The computer software and Arduino remain in communication to allow further user interaction. Other settings, such as the desired LCVR voltages, temperature, camera readout patterns and gain, are sent directly to the devices over USB cables.

\section{Differential data analysis}
\label{datareduction}
They key to VAMPIRES' performance is its differential measurement process, which is based on a calibration strategy adapted from conventional masking interferometry (see Section~\ref{NRM}). However, rather than employing a separate PSF reference star, calibration is performed between Fourier observables extracted from images in simultaneously-recorded orthogonal polarisations. This differential multi-tiered calibration process removes most sources of spatially and temporally-dependent systematic error, producing a purely polarimetric set of observables. 

Because calibration takes place on Fourier domain observables, the first step is the extraction of the two interferograms arising from each Wollaston channel in every camera frame to be windowed and Fourier transformed. The visibilities are then extracted from the power spectrum, and accumulated over a data cube (corresponding to a given LCVR and HWP state) consisting of around 200 frames. The bispectrum (the triple-product of the complex amplitudes of three baselines forming a closing triangle) is also accumulated, and the argument of the accumulated bispectrum gives the closure phase. For the non-redundant aperture masks, the complex visibility data are extracted at the $uv$ coordinates corresponding to the set of known baselines formed by the mask. For the partially-redundant annulus, discreet baselines are not present, so instead the Fourier domain is sampled uniformly while avoiding the regions of low power associated with gaps in the annulus (covering the secondary-mirror support structures as depicted in Figure~\ref{fig_apmasks}).

\begin{figure}
    \centering
    \includegraphics[width=0.48\textwidth]{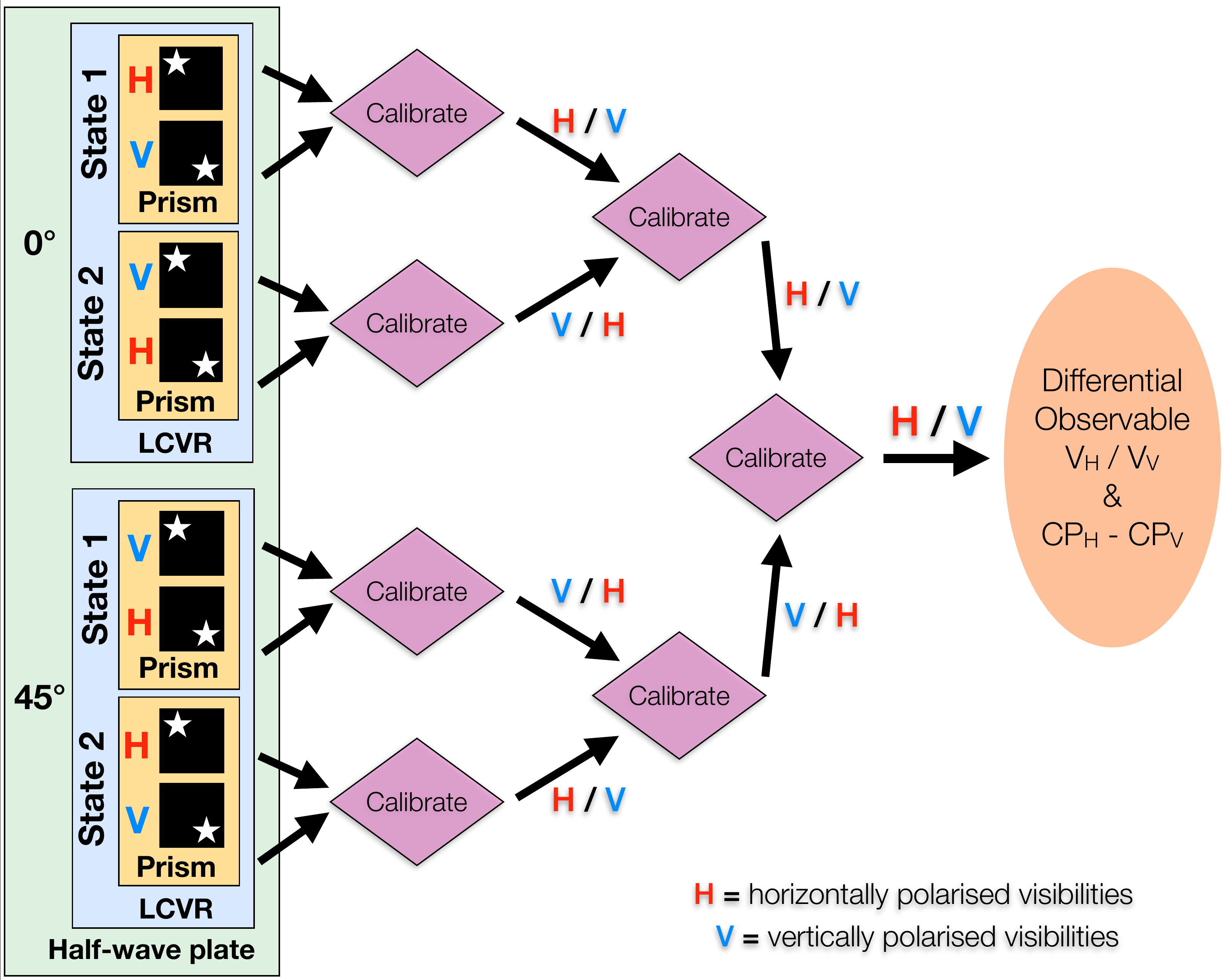}
    \caption{A pictorial depiction of our three-tiered calibration procedure. Calibration is performed between orthogonal polarisations rather than between a target and PSF calibrator star, to yield a polarimetric differential observable. The star in the black squares represents the interferogram as it appears on the detector. The `H' or `V' next to the black square indicates whether this polarisation is Horizontal or Vertical, as determined by the combination of LCVR state, HWP position and Wollaston prism channel. The top-most image is denoted `horizontal' for purposes of demonstration. A series of divisions of visibilities create the final differential observable whilst systematic errors are cancelled out; an analogous process is performed with the closure phases using differences rather than ratios.}
    \label{fig_caldiag}
\end{figure}

An overview of the calibration process is depicted in Figure~\ref{fig_caldiag}. The Wollaston prism allows measurements of orthogonal polarisations to be taken simultaneously and calibrated against each other, resistant to time-varying errors but subject to non-common path error. Conversely, the fast channel-switching LCVR allows the two channels of the Wollaston prism to be switched, and calibration performed between channel-switched states. This removes the effect of non-common path in the Wollaston, although with switching timescales longer than $\tau_0$ (the atmospheric coherence time, due to seeing) it is subject to some time-varying error. The calibration of these two calibrated quantities against one another - forming a second tier of calibration - provides resistance against both these error types. Finally, channel-switching of the bulk of the instrument takes place via a rotating half-wave plate upstream. This calibrates out spatially-dependent systematic errors due to the intervening optics. The rotating HWP also allows both linear Stokes parameters (Q and U) to be measured by rotating the polarisation 45 degrees.

For example, the visibilities from the two channels of the Wollaston prism ($V_{\rm Ch1}$, $V_{\rm Ch2}$) may be calibrated against one another to produce 
\begin{align}
    \frac{V_{\rm Ch1}}{V_{\rm Ch2}} = \frac{V_{\rm Horiz}}{V_{\rm Vert}}
\end{align}
This is repeated but with polarisations rotated by 90 degrees using the LCVR, yielding 
\begin{align}
    \frac{V_{\rm Ch1}'}{V_{\rm Ch2}'} = \frac{V_{\rm Vert}}{V_{\rm Horiz}}
\end{align}
These two differential quantities are then calibrated against one another, and raised to the power of $1/2$ to maintain units:
\begin{align}
    \left( \frac{V_{\rm Ch1}}{V_{\rm Ch2}} / \frac{V_{\rm Ch1}'}{V_{\rm Ch2}'} \right)^\frac{1}{2} &= \left( \frac{V_{\rm Horiz}}{V_{\rm Vert}} / \frac{V_{\rm Vert}}{V_{\rm Horiz}} \right)^\frac{1}{2} \\
    &= \left( \frac{V_{\rm Horiz}^2}{V_{\rm Vert}^2} \right) ^\frac{1}{2} \\
    &= \frac{V_{\rm Horiz}}{V_{\rm Vert}}
\end{align}
This is repeated for all other states and half-wave plate positions, as depicted in Figure~\ref{fig_caldiag}, resulting in a final, calibrated $V_{\rm Horiz}/V_{\rm Vert}$ observable cleaned of almost all temporal and non-common-path errors. The same process takes place for the closure phases, but with differences rather than ratios, to produce a final calibrated $CP_{\rm Horiz}-CP_{\rm Vert}$. Note that the `horizontal' and `vertical' polarisation angles referred to here are arbitrary (ultimately mapping to a known position angle on sky), and conceptually represent any two orthogonal states. The entire process is repeated for both Stokes Q and U by offsetting the angle of the HWP.

\section{On-sky results}
\label{onsky}
The VAMPIRES instrument was tested on-sky in July 2013 during SCExAO engineering time which allowed the calibration precision to be measured under real observing conditions. These tests were conducted with the standard AO\,188 system. The system configuration at the time had relatively low throughput, with the visible beam being split between several different instruments both spectrally and with a 50/50 grey beam-splitter. Therefore we expect the precisions shown here to improve in subsequent observations. The goal is to demonstrate differential visibility precision of order $10^{-3}$ (or 0.1\%).

\subsection{On-sky differential visibilities and closure-phases}
\label{onskydiff}
Two unresolved bright stars, Vega and Altair, were observed as part of the SCExAO engineering schedule, while the VAMPIRES instrument was able to obtain simultaneous observations in the visible.
This provided an ideal test of VAMPIRES' calibration precision, since these point-source\footnote{The known disk around Vega is significantly too large to be seen with VAMPIRES.} stars should ideally exhibit a polarised-differential visibility of 1.0 on all baselines, and zero closure phases. This also allowed the evaluation of VAMPIRES' multiple tiers of differential calibration, both individually and in combination. 

\begin{figure}
    \centering
    \includegraphics[width=0.5\textwidth]{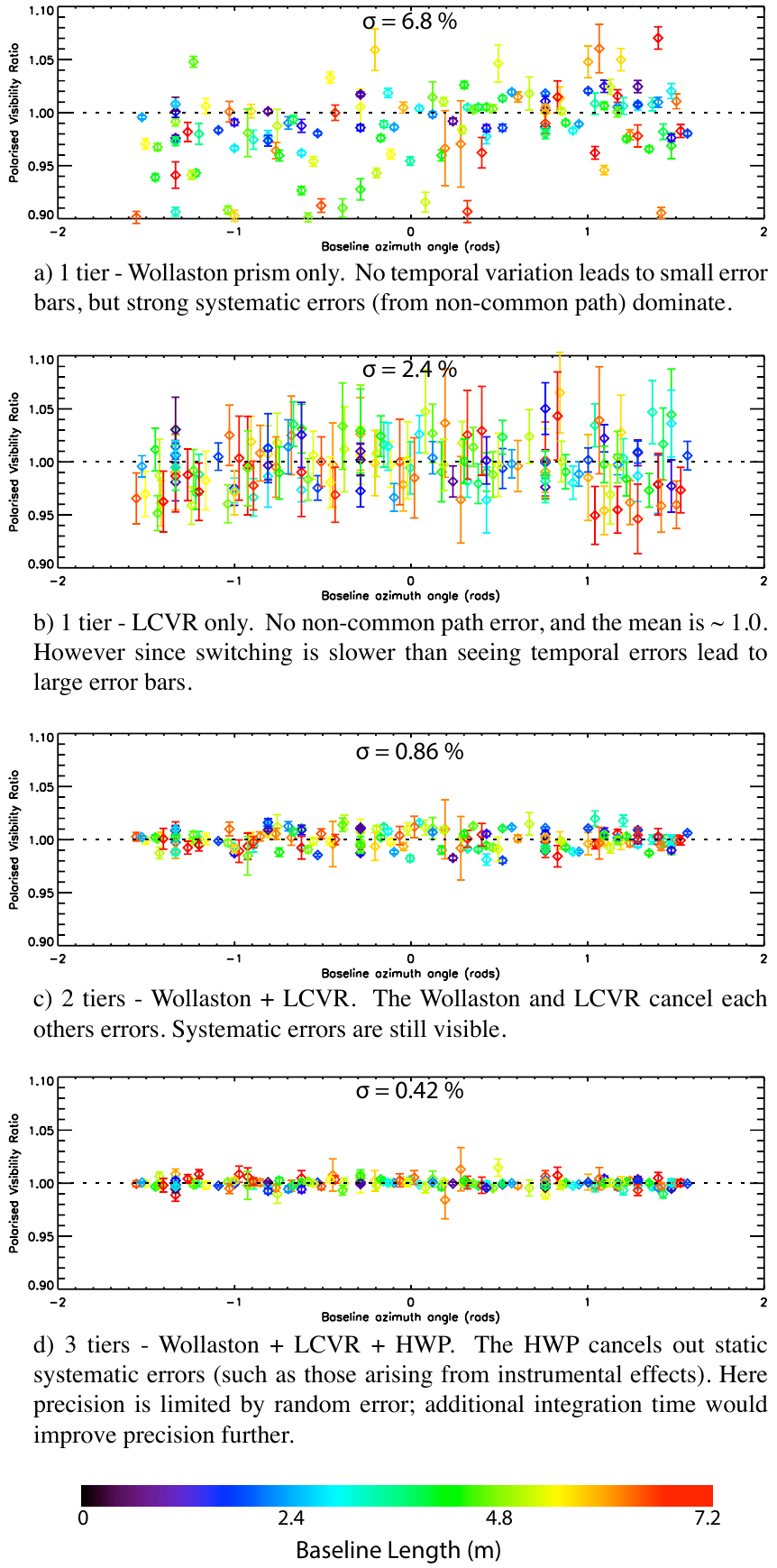}
    \caption{The on-sky differential visibilities from an observation of Vega at 775~nm with the 18 hole mask, showing the effect of different tiers of calibration. Ideally the visibility ratio should be unity on all baselines, since the source is unresolved. Baseline azimuth is plotted on the horizontal axis, while baseline length is represented by colour. The precision is seen to increase with successive layers of calibration, as discussed in the text. Data were taken without Extreme-AO correction.}
    \label{vega_vhvv}
\end{figure}

The results of observing Vega with the 18 hole mask at $\lambda$ = 775~nm (FWHM = 50~nm) with a total integration time of 109~s are shown in Figure~\ref{vega_vhvv}. The standard deviation of the differential visibilities is seen to increase with successive tiers of differential calibration. The Wollaston prism in (a) removes time-varying errors (e.g. due to seeing) so small random-errors are reflected in the error-bars, but non-common path error leads to large (up to 10\%) systematics. Conversely, the LCVR (b) removes non-common path error but is subject to time-varying error, resulting in a mean of $\sim$1 but large random error. Double-differential calibration (c) with the LCVR and Wollaston mitigates both time-varying and non-common path error. Triple-differential calibration (d) with the half-wave plate further removes static systematic errors (such as those arising from instrumental effects, and non-uniform retardance across the aperture of the LCVR due to thermal gradients). The resulting differential visibilities have a standard deviation of $4.2 \times 10^{-3}$, which is approaching our desired performance levels. At this point the precision is limited by random noise processes (such as photon and EM gain noise) and would improve with longer integrations. It should also be noted that these data were taken without the Extreme-AO correction anticipated from SCExAO in the future, which will lead to further improvement (increased fringe visibility and hence increased signal/noise). A similar observational sequence on Altair was conducted with the 9~hole mask (which trades more sparse Fourier sampling for the gain of higher throughput) at $\lambda$ = 750~nm (FWHM = 40~nm). Here the triple-differential visibilities had a standard deviation of $2.4 \times 10^{-3}$.

\begin{figure}
    \centering
    \includegraphics[width=0.5\textwidth]{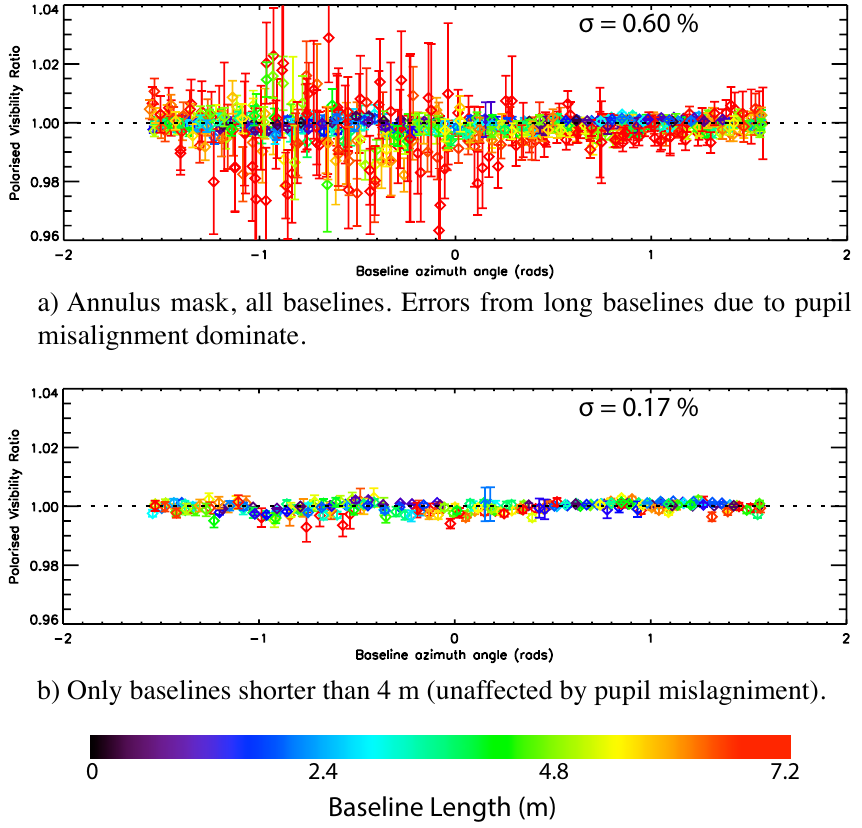}
    \caption{The on-sky triple differential visibilities from Vega at 775~nm, with the annulus mask. Due to a misalignment between the mask and pupil, many longer baselines have extremely low visibilities, resulting in large errors (panel a). If these affected baselines are eliminated by only plotting shorter baselines, excellent precision (0.17 \%) is observed (panel b).}
    \label{vega_ann_vhvv}
\end{figure}

The annulus mask was also tested while observing Vega at $\lambda$ = 775~nm, but performed poorly due to a misalignment between the mask and telescope pupil (due to chromatic effects upstream). This mask is especially susceptible to such misalignments since the annulus is oversized, relying on the edge of the telescope pupil to define the outer edge of the masked aperture. One side of the annular opening was almost completely occluded, causing large errors at longer baselines (see Figure~\ref{vega_ann_vhvv}). Performing the same analysis considering only baselines with lengths shorter than 4\,m, then the performance dramatically improves with the standard deviation of the the triple-differential visibilities dropping to $1.7 \times 10^{-3}$ (see Figure~\ref{vega_ann_vhvv}b). VAMPIRES is being upgraded to allow complete remote mask positioning and pupil viewing, eliminating the potential for such a misalignment problem in future observations.

Multiple differential calibration was found to provide improvements in closure-phase precision similar to those described for visibilities. For an unresolved star, the closure phases on all triangles should be zero. Therefore for the ensemble of all closure phases should ideally have a mean of zero and standard deviation of zero. 

\begin{figure}
    \centering
    \includegraphics[width=0.5\textwidth]{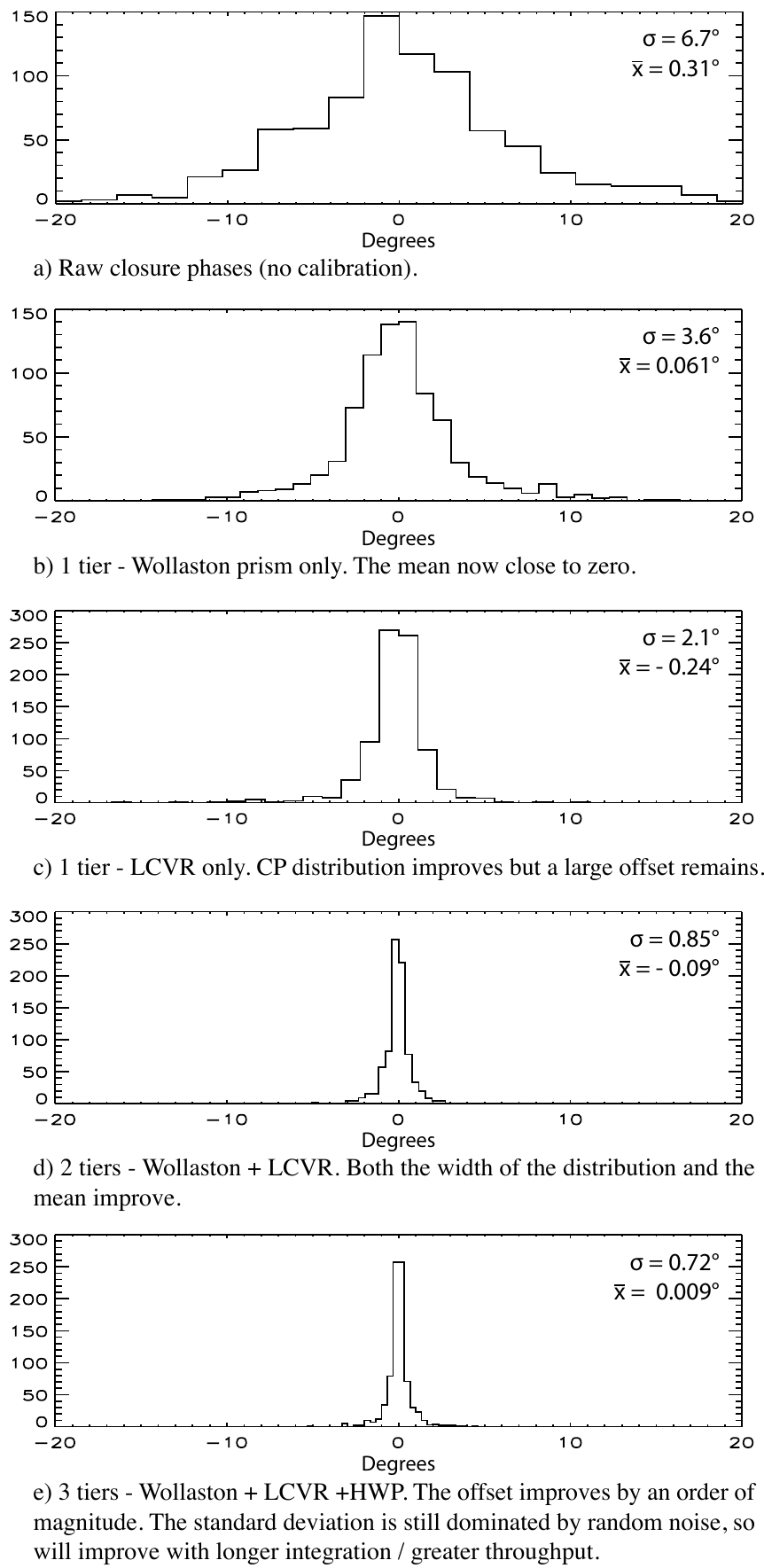}
    \caption{Histograms of the on-sky differential closure phases recorded on Vega (same data as presented in Figure~\ref{vega_vhvv}), showing the effect of different calibration levels. Since the star is unresolved, ideally both the mean and standard deviation would be zero.}
    \label{vega_diffcp}
\end{figure}

A set of histograms of closure phases (for the same Vega observation as in Figure~\ref{vega_vhvv}) at varying levels of differential calibration are given in Figure~\ref{vega_diffcp}. As found for the visibilities, the precision -- represented by a small mean and width of the distribution -- improves with increasing levels of calibration. With the full triple-calibration applied, the closure phase distribution has a standard deviation of 0.72$^\circ$ and a mean of 0.009$^\circ$. However examination of the random error for each triangle (based on the error encountered across the set of all integrations) is large compared to the standard error of the ensemble of closure phases, indicating that statistical errors set the present limitation and precision will improve further with larger volumes of data.

\subsection{Non-polarimetric on-sky results}
\label{non-pol-nrm}
As previously mentioned, VAMPIRES can also operate in a non-polarimetric mode, wherein it works the same way as conventional aperture masking (albeit at shorter wavelengths). In this case, signals from orthogonal polarisation channels are simply combined together, thus discarding the polarisation information. Calibration was now performed with respect to a separate observation of an unresolved reference star. During the July 2013 observations, VAMPIRES had the opportunity to observe $\chi$~Cygni, an S-type star expected to be spatially resolved, and the binary star $\eta$~Pegasi.

$\chi$~Cygni was observed by VAMPIRES with the 18 hole mask at $\lambda$ = 750 nm, and was resolved, allowing its angular diameter to be measured. The V magnitude was $\sim$6 at time of observation\footnote{Observations from the AAVSO International Database, http://www.aavso.org}, and total integration time was 54~s. No polarised signal was detected from this target during these observations within the sensitivity achieved in this observation (differential visibility precision of $\sim$ 2 \% due to the relatively short integration time and faintness of the target). However non-polarimetric measurements were made using the previously observed star Altair (discussed above) as a PSF reference, although this was not an ideal calibrator since it was observed at a different air-mass and time of the night. Despite this, accurate complex visibility data were recovered, constraining a uniform disk fit yielding a diameter of 32.2~$\pm$~0.1 milliarcseconds. This is in close agreement with the literature values tabulated in the CHARM2 catalog \citep{Richichi2005}, which gives the uniform disk diameter as 32.8 $\pm$ 4.1 milliarcseconds in V band.

The binary system $\eta$~Pegasi was observed with the 18 hole mask at $\lambda$ = 775 nm, again for a total integration time of 54~s. Vega was again used as a calibrator (with the same reservations). The binary was detected, and its separation and position angle constrained. A Monte Carlo simulation was used to determine the statistical confidence of the detection, which was found to be better than 99.9\%. The separation was measured to be 48.9~$\pm$~0.6\,mas. This is consistent with the predicted separation based on the orbital parameters measured by \cite{Hummel1998} of 49.9\,mas. The slight discrepancy is probably a result of imperfect knowledge of the mapping between the sky and instrumental field orientations, which is presently based only on values from the optical system model.
Further studies of several stellar systems with known structure are planned to precisely calibrate both orientation and plate scale of VAMPIRES. 
The contrast ratio was measured to be 3.55~$\pm$~0.06 magnitudes, again in good agreement with the value measured by \cite{Hummel1998} of 3.61~$\pm$~0.05 magnitudes.

\section{Simulated data and performance predictions}
\label{sec_simdata}

The differential Fourier visibilities (e.g. $V_{\rm Horiz}/V_{\rm Vert}$) obtained from VAMPIRES are not directly equivalent to the differential intensities (or fractional polarisations) obtained in techniques such as polarised differential imaging. Rather, the magnitude of each of these differential visibilities (i.e. of their departure from $V_{\rm Horiz}/V_{\rm Vert} = 1$) describes the amount of correlated polarised flux at the corresponding spatial frequency, and the differential closure phases describe the corresponding phase. Since these quantities are less intuitive than polarised intensities, simulated VAMPIRES data has been produced using a radiative transfer model of a representative flared axisymmetric protoplanetary disk, in order to give a qualitative example. This particular model demonstrates the ability of VAMPIRES to precisely observe the inner rim of such disks. The model was created using the Hyperion radiative transfer code \citep{Robitaille2011} using a parametric density function, with a power-law surface density profile and Gaussian vertical structure \citep{Andrews2011}. The disk density is given by
\begin{equation}
\rho(R,z,\phi) = \rho_0 \left( \frac{R_0}{R} \right)^{\beta-p} \exp{\left( -\frac{1}{2} \left( \frac{z}{h(R)} \right)^2 \right)}
\end{equation}
where
\begin{equation}
h(R) = h_0 \left( \frac{R}{R_0} \right) ^\beta .
\end{equation}

\begin{figure*}
    \centering
    \includegraphics[width=1.0\textwidth]{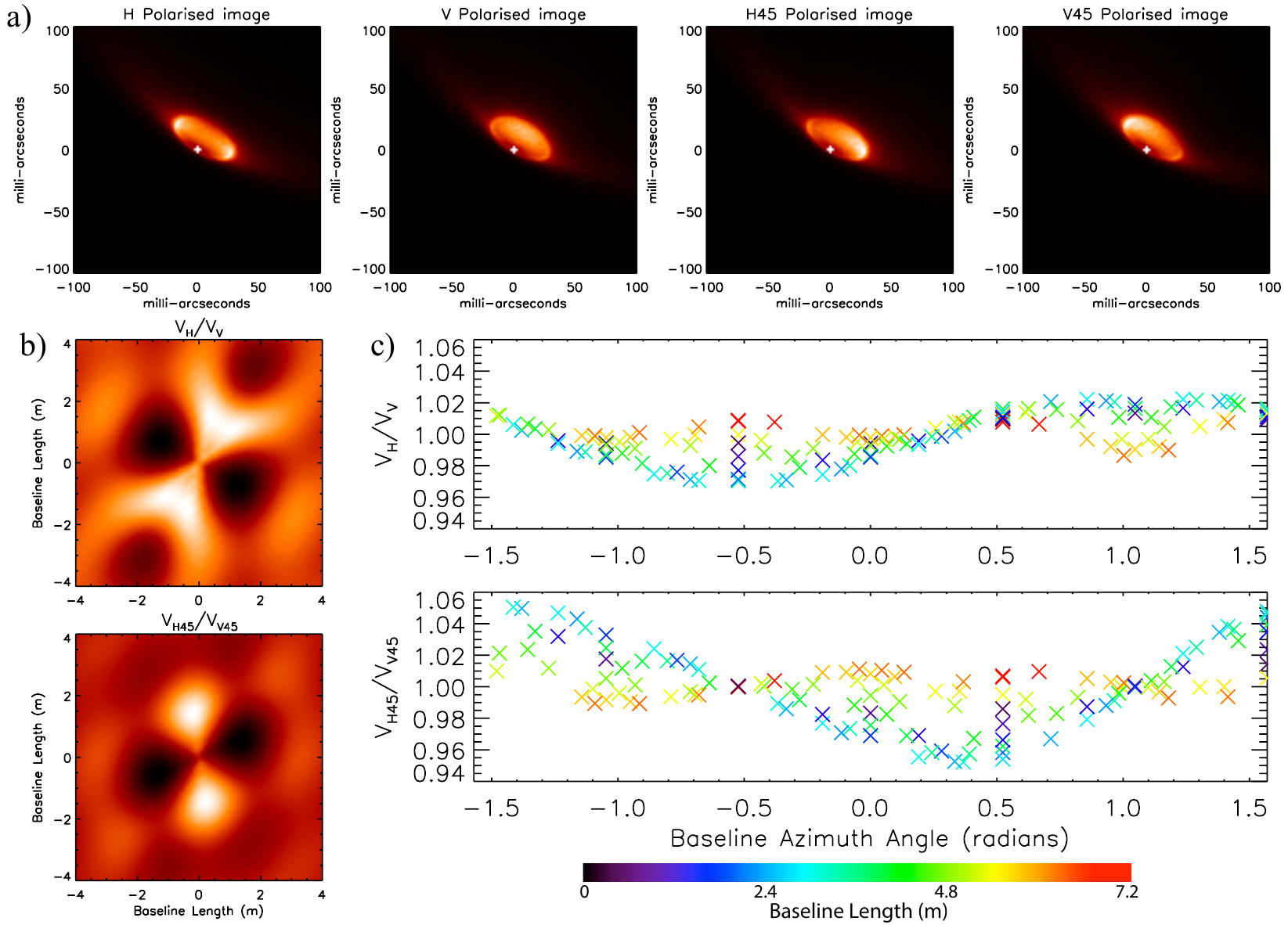}
    \caption{A modelled protoplanetary disk at 500~pc (see Section~\ref{sec_simdata}) and the derived VAMPIRES data for $\lambda$ = 800~nm. a) Image of the inner region of the disk, shown with a non-linear intensity mapping, in four polarisations: horizontal H and vertical V corresponding to Stokes Q, while H45 and V45 are the two orthogonal polarisations rotated 45$^\circ$ corresponding to Stokes U. b) The differential power spectra for the two pairs of orthogonal polarisations. c) Expected differential visibility signals as seen by VAMPIRES with an 18 hole mask.}
    \label{fig_simdata}
\end{figure*}

Parameters were set to typical values, with an inner radius of 25~au and an outer radius of 300~au, and a disk mass of 0.01 M$_\odot$, from which $\rho_0$ is calculated automatically. The surface density exponent $p$ was set to -1 and the scale-height exponent $\beta$ set to 1.125. $H_{100}$, the scale-height of the disk at 100~au, was set to 20~au. The inclination and position angle are 57$^\circ$ and 30$^\circ$ respectively. The model disk surrounds an A0 type star with $T_{eff}$ = 9000~K. The dust in the disk is a mix of silicate and carbon species based on the KMH distribution \citep{Kim1994}. The observational wavelength is 800~nm and the disk is placed at a distance of 500~pc.

polarised images produced by the Hyperion code, together with the corresponding polarised differential visibilities, are shown in Figure~\ref{fig_simdata}. In this model scenario starlight scattered by the inner wall dominates, and the VAMPIRES data is seen to be extremely sensitive to the structure of the inner region. A strong modulation of the differential visibilities as a function of azimuth and baseline length is apparent in Figure~\ref{fig_simdata}c. Along with the associated differential closure phases (not plotted here), these encode the detailed polarised structure of the inner 100~mas or so. If more complex structures were present, such as asymmetrical clumps or an inner disk, they would be clearly evident in these signals.

The on-sky sensitivity of VAMPIRES has been quantified based on laboratory testing, wherein flux levels matching various stellar magnitudes were used along with simulated atmospheric turbulence. The turbulence was simulated by reproducing a AO-corrected Kolmogorov screen (based on typical seeing and wind speeds) via active modulation of the wavefront using the 2K deformable mirror within SCExAO. Visibility precisions of order $10^{-3}$ were reliably obtained for targets as faint as 4th magnitude using the 18 hole mask for integration times of 15 minutes (plus overhead for waveplate switching), or 5.5 magnitudes in 1 hour integration. 
For fainter targets a mask with higher throughput is advantageous. 
The laboratory test data indicated that the 9 hole mask will achieve this precision in 1 hour for 6.5 magnitude stars and the 7 hole mask will achieve it in 1 hour for 8th magnitude. 
The annulus should theoretically achieve order $10^{-3}$ precision in 1 hour integration for stars as faint as 8.5 magnitude, however as discussed in Section \ref{onskydiff} this is currently only realised on the shorter ($<$~4~m) baselines due to pupil alignment drift, a problem that is currently being addressed. Additional overhead must be allowed for due to time taken for waveplate switching - this can be anywhere between 25\% and 100\% of the integration time depending on switching frequency.

The representative model presented in this section has differential visibilities with an average magnitude (deviation for unity) of approximately 2\%. With the demonstrated on-sky precision using the 18 hole mask of 0.4\%, the $V_{\rm Horiz}/V_{\rm Vert}$ of each baseline can be measured to 5$\sigma$. However the actual uncertainties on a fitted model would be much smaller due to the relatively small number of free parameters involved.

\section{Summary}
\label{Summary}
By combining non-redundant aperture-masking interferometry with differential polarimetry, the VAMPIRES instrument will directly image the inner-most region of protoplanetary disks, providing critical insight into the processes of disk evolution and planet formation. Non-redundant aperture masking provides diffraction-limited performance by way of the established interferometric visibility and closure-phase observables. VAMPIRES' triple-differential polarimetric calibration strategy exploits the polarisation of scattered starlight, utilising simultaneous differential measurements with a Wollaston prism, fast channel-switching with a liquid-crystal variable retarder and slow-switching with a rotating half-wave plate to better remove instrumental systematics. The resulting signal encodes the resolved, polarised structure of the inner disk. These observables are largely immune to the effects of instrumental polarisation, with the remainder being removed by precise calibration of the instrumental Mueller matrix using an in-built characterisation system.

VAMPIRES records data at visible wavelengths (where polarisation from scattering is typically higher) in a hitch-hiker mode that does not affect simultaneous science operation of other (infrared) instruments. On-sky demonstrations of the VAMPIRES instrument yielded a differential-visibility precision approaching $10^{-3}$ and closure phase standard-deviation better than $1^\circ$. Limitations to both performance metrics are presently provided by restricted statistical sample size and therefore further improvement is expected with longer on-sky integration times. Precise visibilities and closure phases will be used to accurately constrain disk models, and to detect the presence of asymmetries and density enhancements which reveal the presence of embedded gravitating companions. The instrument is now integrated into the SCExAO system and is ready for its first science observations, planned for mid to late 2014. Eventually the instrument will be largely autonomous and capable of entirely remote operation, allowing simultaneous measurements with standard facility instruments SCExAO/HICIAO when required.

\bibliographystyle{mn2e}
\bibliography{VAMPIRES2014}

\appendix
\section{Calibration of instrumental polarisation}
\label{IPCal}
The differential calibration of VAMPIRES mitigates the effect of many instrumental polarisation effects. Moreover, since the visibilities for each polarisation channel are normalised with respect to the total flux in that channel, simple diattenuation has no effect on the measured observables (unlike the case for standard polarimetry). However this does not take into account cross-terms in the instrumental Mueller matrix (which describes the polarisation properties intrinsic to the instrument and how they alter each of the Stokes parameters of the incoming light \citep{Goldstein2011}). Q~$\rightarrow$~U and U~$\rightarrow$~Q mixing will lead to incorrect measurement of the relative magnitude of Stokes Q and U measurements. The Q~$\leftrightarrow$~V and U~$\leftrightarrow$~V cross-terms are even more problematic - since VAMPIRES does not measure Stokes V, it thus appears to the instrument that part of the Stokes Q and U components have `disappeared', leading to an underestimation of the magnitude of the polarisation. Since VAMPIRES itself is behind the pre-existing systems AO\,188  (which includes a k-mirror image rotator) and SCExAO, significant instrumental polarisation cannot be avoided, and instead must be mitigated by careful calibration and the tripe-differential measurement process. 

To resolve these problems, a careful full characterisation of the instrumental Mueller matrix is performed immediately before or after astronomical observations, which is then used to correct data during processing. This characterisation procedure is fully automated and can be performed remotely. First, a linear polariser on a rotation stage is driven into the beam upstream of the SCExAO optical path, and a rotating quarter-wave plate is positioned immediately before the Wollaston prism inside VAMPIRES - see Figure~\ref{fig_schematic}. Any residual polarised structure in the light incident upon the linear polariser is removed by first passing it (as a large-diameter beam) through an achromatic wedge depolariser, allowing an arbitrary linear polarisation to be generated. Alternatively, the halogen flat-field lamp, linear polariser and half-wave plate which already exist within the AO188 adaptive optics system may be used to inject the linearly polarised reference, which has the advantage of characterising the optics within AO188 itself as well. By using the Wollaston prism as the analyser, a rotating-polariser / rotating-compensator+fixed-analyser (RP/RCFA) type Mueller matrix polarimeter \citep{Hauge1980} is created. Data from such a setup can specify the first three columns of the Mueller matrix of the instrument. While the fourth column can not be determined, if the assumption is made that the astrophysical source has a negligible circular polarisation component, then the missing fourth column is inconsequential. 

If the quarter-wave plate and linear polariser are rotated synchronously such that the angle of the quarter-wave plate is three times that of the polariser, then the first three columns of the Mueller matrix can be directly determined by Fourier analysis of the resulting intensity variation \citep{Hauge1980}. Alternatively, to provide more physical insight into the origin of the polarisation effects, a polarisation model of the instrument can be created, and then fine-tuned by fitting it to the measured calibration data \citep{Witzel2011}. In this case, the Mueller matrix of the instrument is created by combining the Mueller matrices of the individual components, with the appropriate rotations. The linear polariser (and the Wollaston prism channels), with their polarisation axis at angle $\theta$, are represented by the matrix $M_{\rm LP}$:
\begin{equation}
M_{\rm LP} = M_{r(\theta)}^{-1} \times M_{LP(h)} \times M_{r(\theta)}
\end{equation}
where $\times$ signifies matrix multiplication, $M_{LP(h)}$ represents the matrix of an ideal horizontal linear polariser, i.e.:
\begin{equation}
\label{lpideal}
M_{LP(h)} = \frac{1}{2}
\begin{pmatrix}
    1 & 1 & 0     & 0     \\
    1 & 1 & 0     & 0     \\
    0           & 0           & 0 & 0  \\
    0           & 0           & 0        & 0
\end{pmatrix}
\end{equation}
and $M_{r(\theta)}$ is the rotation matrix in Stokes space for angle $\theta$:
\begin{equation}
\label{rotmat}
M_{r(\theta)} = 
\begin{pmatrix}
    1 & 0 & 0     & 0     \\
    0 & \cos{2\theta} & \sin{2\theta}     & 0     \\
    0 & -\sin{2\theta}  &  \cos{2\theta} &  0 \\
    0           & 0           & 0        & 1
\end{pmatrix}.
\end{equation}

Similarly, the Mueller matrix for a wave-plate (retarder) can be represented as 
\begin{equation}
M_{\rm WP} = M_{r(\theta)}^{-1} \times M_{WP(h)} \times M_{r(\theta)}
\end{equation}
where $M_{r(\theta)}$ is as before and $M_{WP(h)}$ is the matrix of a wave-plate with retardance $\phi$:
\begin{equation}
M_{WP(h)} =
\begin{pmatrix}
    1 & 0 & 0     & 0     \\
    0 & 1 & 0     & 0     \\
    0           & 0           & \cos{\phi} & \sin{\phi}  \\
    0           & 0           & -\sin{\phi}        & \cos{\phi}
\end{pmatrix}
\end{equation}

Combinations of these matrices can represent VAMPIRES' half-wave plate, LCVR and Wollaston prism, as well as the linear polariser and quarter-wave plate used for calibration. Instrumental polarisation arises mostly from reflections off various mirrors in the system. Reflections from a metallic mirror cause both a change in transmission between linear polarisation components and a change in phase between these components. The Mueller matrix for a metallic mirror can thus be constructed by combining the matrices for a partial-linear polariser with a wave plate, resulting in the following matrix \citep{Clarke1973}: 
\begin{equation}
M = \frac{1}{2}
\begin{pmatrix}
    r_\perp + r_\parallel & r_\perp - r_\parallel & 0     & 0     \\
    r_\perp - r_\parallel & r_\perp + r_\parallel & 0     & 0     \\
    0           & 0           & \sqrt{r_\perp r_\parallel}\cos{\delta} & \sqrt{r_\perp r_\parallel}\sin{\delta}  \\
    0           & 0           & -\sqrt{r_\perp r_\parallel}\sin{\delta} & \sqrt{r_\perp r_\parallel}\cos{\delta}
\end{pmatrix}
\end{equation}
where $r_\perp$ and $r_\parallel$ are the coefficients of reflection for the perpendicular and parallel polarisations respectively, and $\delta$ is the retardance between the components. For each metallic surface in the VAMPIRES optical model, these values are in turn calculated from the metal's known complex refractive index $\widetilde{n_2}$ using the amplitude Fresnel equations (where $\widetilde{n_1} \approx 1$ is the refractive index of air). The complex amplitudes of the perpendicular and parallel reflected components, respectively, are given by
\begin{align}
\widetilde{r_\perp} &= \frac{\widetilde{n_1}\cos{\theta_i} - \widetilde{n_2}\cos{\theta_t}}{\widetilde{n_1}\cos{\theta_i} + \widetilde{n_2}\cos{\theta_t}}
\end{align}
\begin{align}
\widetilde{r_\parallel} &= \frac{\widetilde{n_2}\cos{\theta_i} - \widetilde{n_1}\cos{\theta_t}}{\widetilde{n_1}\cos{\theta_t} + \widetilde{n_2}\cos{\theta_i}}
\end{align}
where $\theta_i$ is the angle of incidence and $\theta_t$ is nominally the angle of transmission, and is calculated using Snell's law, but in the case of a metallic reflection it is complex. The coefficients of reflection are then simply
\begin{align}
r_\perp = |\widetilde{r_\perp}|^2, \quad r_\parallel = |\widetilde{r_\parallel}|^2
\end{align}
and the relative retardance is just
\begin{align}
\delta = \arg{(\widetilde{r_\perp})} - \arg{(\widetilde{r_\parallel})} .
\end{align}

A Mueller matrix for the entire instrument is thus created by combining the matrices of all polarising elements, with the appropriate rotations. Using the Levenberg-Marquardt algorithm (or a simple parameter grid), a fit of the model to the measured calibration data is performed, in which the free parameters are the complex refractive indices of the metallic surfaces (starting at the tabulated value\footnote{Complex refractive indices obtained from http://refractiveindex.info/. Optics suppliers unfortunately do not provide this level of characterisation.} for the specific metal comprising the mirror coating). 
This fitting process also allows the dichroic mirror (of poorly-known reflection characteristics) to be characterised. 
To reduce the number of free parameters, the SCExAO focusing mirror and the DM (which are both in the same plane of reflection) are combined into one component.

Finally, the instrumental polarisation contribution from the telescope itself and the AO\,188 system must be taken into account. In principle it should be straightforward to probe this portion of the optical system with observations of polarised standard stars, but until such data can be obtained, detailed optical modelling must serve. Fortunately a precise ZEMAX model of the system was made available to us, yielding precise knowledge of position and angle of all optical surfaces. Alternatively, the contribution from AO\,188 can be measured by using its own flat-field lamp, linear polariser and half-wave plate. Exploiting this, an accurate matrix $M_{tel}$ can be calculated using the methods discussed above.

Thus a final Mueller matrix for the instrument is constructed:
\begin{equation}
\begin{split}
M_{\rm VAMPIRES} = &M_{\rm Woll} \times M_{\rm QWP} \times M_{\rm LCVR} \times M_{\rm PerM2} \\
&\times M_{r(\theta_{Per})} \times M_{\rm HWP} \times M_{\rm PerM1} \\
&\times M_{\rm OAP+DM} \times M_{\rm LP} \times M_{tel}
\end{split}
\end{equation}
where each matrix term incorporates its appropriate rotation matrix. $M_{\rm OAP+DM}$ represents the combined in-plane mirrors on the bottom bench, $M_{\rm PerM1~\&~2}$ are the periscope mirrors, $M_{r(\theta_{per})}$ is the beam rotation of the periscope and $M_{\rm Woll}$ the Wollaston prism, which in fact exists in two instances (with $\pm 45^\circ$ rotations corresponding to the two channels). The calibration polariser and wave-plate ($ M_{\rm QWP}$ and $M_{\rm LP}$) are included for fitting the model to the calibration data. An alternative configuration under investigation replaces the HWP with a pair of quarter-wave plates, which allows the polarisation to be rotated (as with the HWP) but also the system birefringence to be compensated for (by differential rotation of these wave-plates). In this case the $M_{\rm HWP}$ term above is replaced by two $M_{\rm QWP}$ terms.

A sample Mueller matrix for the instrument is given below - in this case the matrix for the instrument at 775~nm in the dual-QWP configuration, for zero polarisation rotation, Wollaston prism ordinary beam and LCVR retardance set to $\pi/2$.

\begin{equation}
M =
\begin{pmatrix}
    1 & -0.026 & -0.062     & -0.015     \\
    -0.022 & 0.706 & 0.225     & -0.668     \\
    0.059           & 0.130           & -0.973 &  -0.189 \\
    0.026           & -0.693           & 0.045        & -0.717
\end{pmatrix}
\end{equation}

Substantial off-diagonal terms are seen (while noting that this matrix includes the beam rotation by the periscope between benches), and as described in Section \ref{datareduction} the bulk of these effects are mitigated by the triple-differential measurement process and the inherent robustness of interferometry against diattenuation (since the signal in each polarisation channel is normalised with respect to the total flux in that channel). (This full matrix, however, is still used for correction during data reduction to ensure any residual effects are calibrated for.)
The effect of birefringence is most strongly seen in the Q~$\leftrightarrow$~V terms rather than the U~$\leftrightarrow$~V terms, due largely to the rotation of the beam by the periscope. This matrix includes the contribution of AO\,188, which makes a sizeable contribution to the instrumental polarisation due to its image rotator (k-mirror).

Polarised light of Stokes vector $S$ incident on the telescope is transformed by the instrument to emerge as $S'$, where
\begin{equation}
S' = M_{\rm VAMPIRES} \times S .
\end{equation}
To determine the signal measured by the camera (which only measures intensity) we apply the detector operator $D$, which is the row vector $[1,0,0,0]$. The intensity measured is then
\begin{equation}
I = D \times M_{\rm VAMPIRES} \times S .
\end{equation}

To correct for instrumental polarisation in the intensity domain, the inverse instrumental Mueller matrix $M_{\rm VAMPIRES}^{-1}$ could simply be applied to the measured Stokes vector. However VAMPIRES' calibration precision relies on immediately transforming each frame of fringes into the Fourier domain and conducting all subsequent operations in this domain, making such a strategy impractical. Therefore rather than applying the instrumental polarisation correction to the data, instead we apply $M_{\rm VAMPIRES}$ directly to the astrophysical model (e.g. a radiative transfer model) or image reconstruction before fitting to the data, which is the technique we will employ in future science observations.

\bsp

\label{lastpage}

\end{document}